\documentclass[12pt]{article}
\usepackage{epsf}
\usepackage{graphicx}

\begin{document}
\title
{Small effects of low-energy quantum gravity}
\author
{Michael A. Ivanov \\
Physics Dept.,\\
Belarus State University of Informatics and Radioelectronics, \\
6 P. Brovka Street,  BY 220027, Minsk, Republic of Belarus.\\
E-mail: michai@mail.by.}

\maketitle

\begin{abstract}Small effects of quantum gravity on the scale $\sim 10^{-3}
eV$ and their cosmological consequences are discussed and compared
with observations of supernovae 1a, gamma-ray bursts and galaxies.
\end{abstract}
Our knowledge of the nature is restricted for many reasons, but
sometimes we attack it with a view of victors being sure that we
know enough to go ahead namely in the given way. An attempt to
introduce dark energy to rescue the picture of expanding universe
seems to me to be such the case.
\par I would like to show here that small effects of very-low-energy
quantum gravity (on the scale $\sim 10^{-3}eV$) \cite{500} can
give an alternative explanation of supernovae 1a, gamma-ray bursts
and galaxy number counts observations. The new picture has the
very dramatic consequence: nor dark energy nor any expansion of
the universe exist in it.
\par There are two small effects in the sea of  super-strong
interacting gravitons \cite{500}: average energy losses of a
photon due to forehead collisions with gravitons and an additional
relaxation of a photonic flux due to non-forehead collisions of
photons with gravitons. The first effect leads to the geometrical
distance/redshift relation: $r(z)= ln (1+z)\cdot c/H,$ where $H$
is the Hubble constant. The both effects lead to the luminosity
distance/redshift relation: $D_{L}(z)=c/H \cdot \ln(1+z)\cdot
(1+z)^{(1+b)/2},$ where the "constant" $b$ belongs to the range 0
- 2.137 \cite{11} ($b=2.137$ for a very soft radiation, and
$b\rightarrow 0$ for a very hard one). For an arbitrary source
spectrum, a value of the factor $b$ should be still computed. It
is clear that in a general case it should depend on a rest-frame
spectrum and on a redshift. Because of this, the Hubble diagram
should be a multivalued function of a redshift: for a given $z,$
$b$ may have different values for different kinds of sources.
Further more, the Hubble diagram may depend on the used procedure
of observations: different parts of rest-frame spectrum will be
characterized with different values of the parameter $b$.
\par In Figure 2 of my paper \cite{500}, the Hubble diagram
$\mu_{0}(z)$ with $b=2.137$ is shown; observational data (82
points) are taken from Table 5 of \cite{203}. The predictions fit
observations very well for roughly $z < 0.5$. It excludes a need
of any dark energy to explain supernovae dimming. Improved
distances to nearby type Ia supernovae (for the range $z < 0.14$)
can be fitted with the function $\mu_{c}(z)$ for a flat Universe
with the concordance cosmology with $\Omega_{M} = 0.30$ and $w
=-1$ \cite{207}. The difference $\mu_{c}(z)-\mu_{0}(z)$ between
this function and distance moduli in the considered model for
$b=1.52$ has the order of $\pm 0.001$ in the considered range of
redshifts \cite{11}. Results from the ESSENCE Supernova Survey
together with other known supernovae 1a observations in the bigger
redshift range $z<1$ can be best fitted in a frame of the
concordance cosmology in which $\Omega_{M} \simeq 0.27$ and $w
=-1$ \cite{208}; the function $\mu_{c}(z)$ for this case is almost
indistinguishable from distance moduli in the considered model for
$b=1.405:$ the difference is not bigger than $\pm 0.035$ for
redshifts $z<1.$
\par Theoretical distance moduli $\mu_{0}(z) = 5 \log D_{L} + 25$ are
shown in Fig. 1 for $b=2.137$ (solid), $b=1$ (dot) and $b=0$
\begin{figure}[th]
\epsfxsize=12.98cm \centerline{\epsfbox{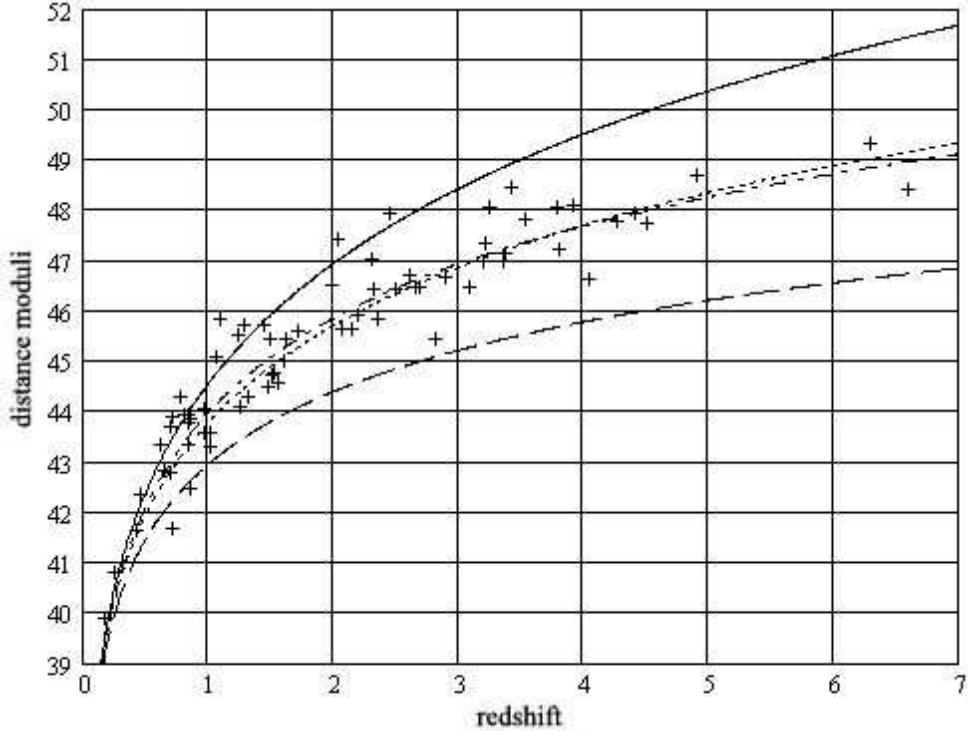}} \caption{
Hubble diagrams $\mu_{0}(z)$ with $b=2.137$ (solid) and $b=0$
(dash); the Hubble diagrams $\mu_{0}(z)$ with $b=1.1$ of this
model (dot) and the one of the concordance model (dadot) which is
the best fit to GRB observations \cite{206}; GRB observational
data (+, 69 points) are taken from Table 6 ($\mu^{a}$) of
\cite{206} by Schaefer.}
\end{figure}
(dash). If this model is true, all observations should lie in the
stripe between lower and upper curves. Theoretical distance moduli
$\mu_{c}(z)$ for a flat Universe with the concordance cosmology
with $\Omega_{M} = 0.27$ and $w =-1$, which give the best fit to
gamma-ray bursts observations \cite{206}, are very close to the
Hubble diagram $\mu_{0}(z)$ with $b=1.1$ of this model. GRB
observational data (+, 69 points) are taken from Table 6
($\mu^{a}$) of \cite{206} by Schaefer.
\par The galaxy number counts/magnitude relation in this model
$f_{3}(m),$ $m$ is a magnitude, in this model (for more detail,
see \cite{12}), which takes into account the Schechter luminosity
function, is based on the same two small effects. To compare this
function with observations by Yasuda et al. \cite{77}, we can
choose the normalizing factor from the condition:
$f_{3}(16)=a(16), $ where $a(m)\equiv A_{\lambda}\cdot
10^{0.6(m-16)}$ is the function giving the best fit to
observations \cite{77}, $A_{\lambda}=const.$ The ratio ${{f_{3}(m)
-a(m)}\over a(m)}$ is shown in Fig. 2 for different values of the
constant $A_{1}\simeq 5\cdot 10^{17}\cdot {L_{\odot} / L_{\ast}}$
by $\alpha =-2.43$ and $b=2.137.$
\begin{figure}[th]
\epsfxsize=12.98cm \centerline{\epsfbox{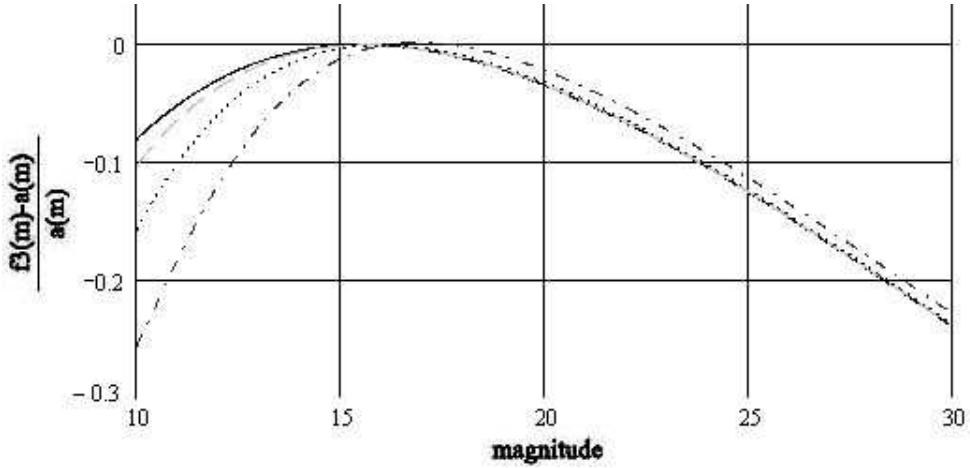}}
\caption{The relative difference $(f_{3}(m)-a(m))/a(m)$ as a
function of the magnitude $m$ for $\alpha=-2.43$ by
$10^{-2}<A_{1}<10^{2}$ (solid), $A_{1}=10^{4}$ (dash),
$A_{1}=10^{5}$ (dot), $A_{1}=10^{6}$ (dadot). }
\end{figure}
If we compare this figure with Figs. 6,10,12 from \cite{77}, we
see that the considered model provides a no-worse fit to galaxy
observations than the function $a(m)$ if the same K-corrections
are added.
\par
\par The considered effects of low-energy quantum gravity are very
small on micro level, but they may be the basic ones for
cosmology. The ones are beyond the general relativity, and
astrophysical observations  seem to stay an unexpected tool of
quantum gravity laboratory.

\end{document}